LA-UR-13-23928
Approved for public release; distribution is unlimited.

Title: Fission Fragments Produced from Proton Irradiation of Thorium Between 40 and 200 MeV

Author(s): Engle, Jonathan
Mashnik, Stepan G.
Weidner, John W.
Fassbender, Michael E.
Bach, Hong
Ullmann, John L.
Couture, Aaron J.
Bitteker, Leo J. Jr.
Gulley, Mark S.
John, Kevin D.
Birnbaum, Eva R.
Nortier, Francois M.

Intended for: 2013 ANS Winter Meeting and Nuclear Expo, 2013-11-10/2013-11-14 (Washington, District Of Columbia, United States)

Issued: 2013-05-30

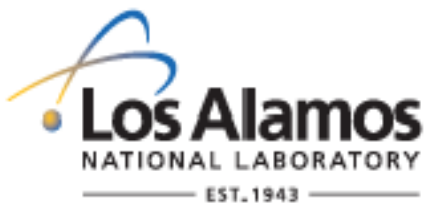

# Fission Fragments Produced from Proton Irradiation of Thorium Between 40 and 200 MeV

Jonathan W. Engle, Stepan G. Mashnik, J. W. Weidner, Michael E. Fassbender, Hong T. Bach, John L. Ullmann, Aaron J. Couture, Leo J. Bitteker, Mark S. Gulley, Kevin D. John, Eva R. Birnbaum, and Francois M. Nortier

[1]Los Alamos National Laboratory, P.O. Box 1663, Los Alamos, NM, 87544, jwengle@lanl.gov

## INTRODUCTION

The cross sections for the formation of five residual radionuclides ($^{72}$Se, $^{97}$Zr, $^{112}$Pd, $^{125}$Sb, and $^{147}$Nb) from 40- to 200-MeV proton irradiation of thorium have been measured and are reported. The atomic masses of these fragments span the expected mass distribution of radionuclides formed by fission of the target nucleus. Especially in mass regions corresponding to transitions between different relaxation mechanisms employed by available models, these data are expected to be useful to the improvement of high-energy transport codes. The predictions of the event generators incorporated into the latest release of the Monte Carlo N-Particle code (MCNP6) are compared with data measured in this work in the hope that these results may be useful to the continued process of code verification and validation in MCNP6.

## EXPERIMENTAL METHODS

Complete discussions of the irradiation methods [1], spectroscopic assay parameters and data analysis techniques [2], and MCNP6 event generators [2] used in this work have been published previously. The following sections describe these experimental details only briefly. These procedures have been successfully used to isolate and quantify the nuclear formation cross sections of the isotopes summarized in Table I below.

Table I. Nuclear data for isotopes measured in this work

| Isotope | $t_{1/2}$ | Decay Mode (*I%*) |
|---|---|---|
| $^{72}$Se | 8.40 d | EC (*100%*) |
| $^{97}$Zr | 16.749 h | β- (*100%*) |
| $^{112}$Pd | 20.03 h | β- (*100%*) |
| $^{125}$Sb | 2.75856 y | β- (*100%*) |
| $^{147}$Nd | 10.98 d | β- (*100%*) |

### Irradiation

Briefly, thorium foils interspersed with aluminum energy degraders, aluminum foils used to monitor proton flux, and stainless steel foils used as beam profile monitors were irradiated with 100- and 200-MeV proton beams in the Target 2 Blue Room of the Los Alamos Weapons Neutron Research Facility. A total of 12 incident proton energies were measured. The $^{27}$Al(p,x)$^{22}$Na cross section was used as a monitor of integrated proton flux. Steel foils were exposed to Gafchromic film in order to confirm the beam's location and profile relative to thorium foils. Irradiated samples were transported to the Nuclear and Radiochemistry Group (C-NR) Countroom, where they were repeatedly assayed by non-destructive high purity germanium (HPGe) gamma spectroscopy for approximately 10 months following each irradiation.

An extensively validated in-house analysis code, SPECANAL, was used to extract photopeak areas from gamma spectra for this work. The activity at the end of bombardment (EoB) of each isotope of interest was determined by fitting of its decay curve, and cross sections were calculated using the well-known activation formula. Uncertainties in linear regressions' fitted parameters were computed from covariance matrices as the standard error in the activity extrapolated to the end of bombardment. This value was combined according to the Gaussian law of error propagation with estimated contributing uncertainties from detector calibration and geometry reproducibility (2.9% combined), target foil dimensions (0.1%), and proton flux (6.8 – 8.1%).

### MCNP6 Event Generators Tested Here

We compare our measured cross sections with predictions of the MCNP6 transport code using three different event generators available in MCNP6 to simulate high-energy nuclear reactions. All predictions were obtained prior to the measurement. A brief summary of the utilized event generators follows; please see ref. [2] and especially works cited therein for a full discussion of these models:

1) The Cascade-Exciton Model (CEM) of nuclear reactions as implemented in the code CEM03.03.

2) The Bertini IntraNuclear Cascade (INC) followed by the Multistage Preequilibrium Model (MPM), followed by the evaporation model as described with the EVAP code by Dresner, followed by or in competition with the RAL fission model (if the atomic number of the compound nucleus Z is ≥ 70), referred to herein simply as "Bertini".

3) The IntraNuclear Cascade model developed at the Liege (INCL) University in Belgium by Prof. Cugnon with his coauthors from CEA, Saclay, France merged with the evaporation-fission model ABLA developed at

GSI, Darmstadt, Germany, referred to herein as "INCL+ABLA".

**Comparison of Experimental and Predicted Data**

These event generators produce several types of data useful to evaluating their predictive power. In particular, each event generator predicts the "independent", or direct, formation of radionuclides from various nuclear reactions. To compare with experimental data it is necessary to take into account the decay of all isotopes contained in the irradiated sample that contribute to the quantity of the isotope of interest by their decay. If no measurements exist during the time frame when a parent's decay contributes to the activity of the daughter, the cross section of the daughter is commonly reported as "cumulative", signifying that the reported value considers parents' decay contributions to the probability that the isotope will be formed as a result of the original physical event (e.g., beam incident on a target). All cross sections reported here are cumulative in nature.

MCNP6 event generators are also capable of reporting cross sections for the production of all isotopes of a given atomic mass number, covering production of isobars that are both neutron-rich and -poor. Plots of these predicted cross sections vs. atomic mass number have the advantage of permitting consideration of larger physical phenomena, in this case fission product distributions produced from a given irradiation, but also of spallation- and evaporation-produced radionuclides whose mass distributions are much heavier (spallation) and lighter (evaporation) than that of fission products. Direct comparison of experimental data with model generated mass distributions would require experimental data from all relevant isobars of the specific mass number being compared, but qualitative comparison is useful in the process of model testing and improvement.

**RESULTS**

Measured cumulative cross sections for the 5 isotopes of interest are compared with the predictions of MCNP6 event generators, which include the formation of all isobars of the same mass number, at three representative energies between 56 and 195 MeV in Fig. 1, parts A - C. As expected, event generator predictions show a broadening of the mass range produced by fission reactions with increasing incident proton energy, and qualitative agreement with the measured data is acceptable for all five isotopes measured. Where possible (at 97 and 195 MeV), plotted measured data also include those measurements reported in [3], whose values are in good agreement with results from the present effort.

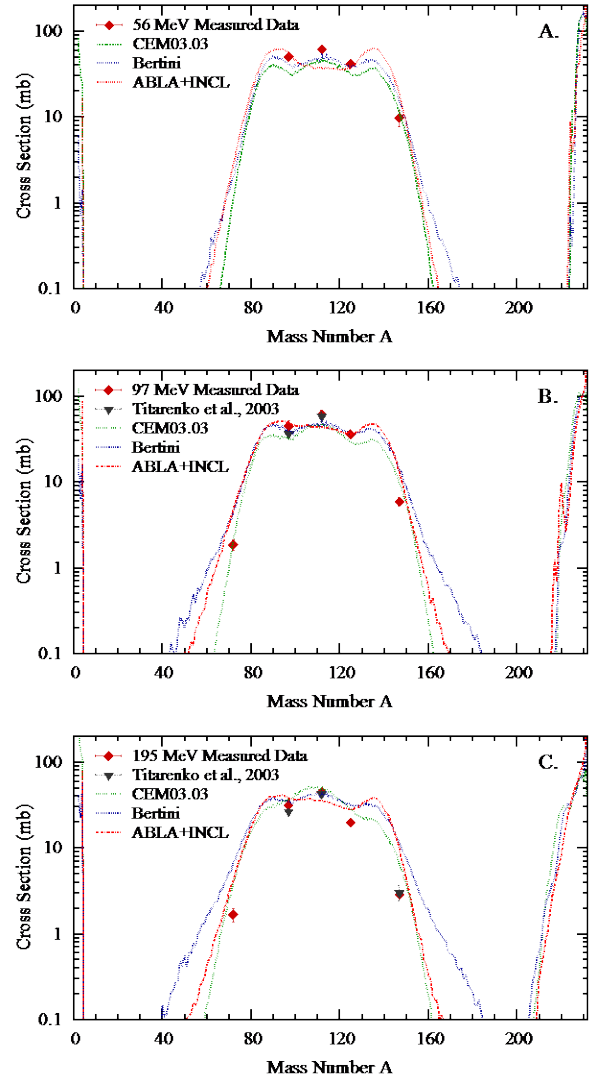

Fig. 1. Qualitative comparison of MCNP6 event generator predicted cross sections for all isobars of the relevant mass number with cumulative cross sections for 5 isotopes measured by the present work at 56 MeV (A), 97 MeV (B), and 195 MeV (C).

The Bertini event generator predicts a broader mass distribution in the fission product spectrum, while CEM03.03's predicted fission product mass distribution is the narrowest of the three models tested. In general, CEM03.03's values are confirmed by measurements at the edges of the fission product mass distribution (near A ~ 45 and A ~ 170) made here and by those reported previously [3]. However, in no case were the cross sections of fission product isobars on opposite sides of "valley of stability" measured by the present work, so direct quantitative comparison between the codes and measured data is impossible per the discussion in the Introduction. The precision of measured data does not permit a more detailed evaluation of the peak structure at the apex of event generators' predicted mass distributions.

The measured nuclear formation cross section is shown for each of the 5 isotopes individually in Figures 2-6.

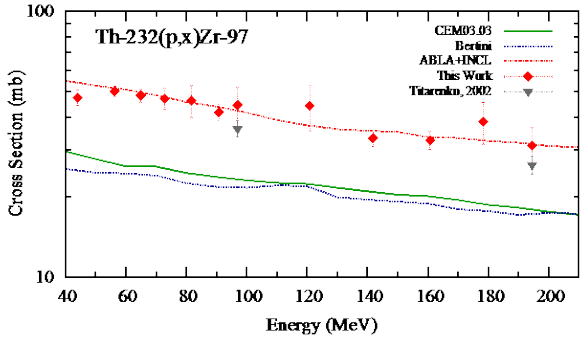

Fig. 2. Comparison of MCNP6 event generator predicted cross sections for the $^{232}$Th(p,x)$^{97}$Zr nuclear reaction from 40- to 200-MeV with experimental data.

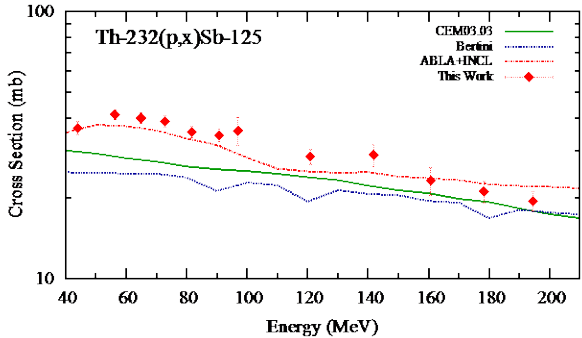

Fig. 3. Comparison of MCNP6 event generator predicted cross sections for the $^{232}$Th(p,x)$^{125}$Sb nuclear reaction from 40- to 200-MeV with experimental data.

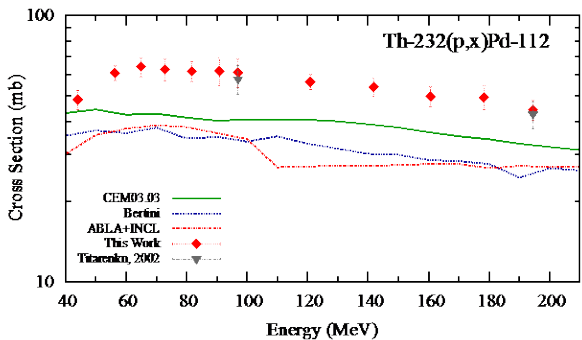

Fig. 4. Comparison of MCNP6 event generator predicted cross sections for the $^{232}$Th(p,x)$^{112}$Pd nuclear reaction from 40- to 200-MeV with experimental data.

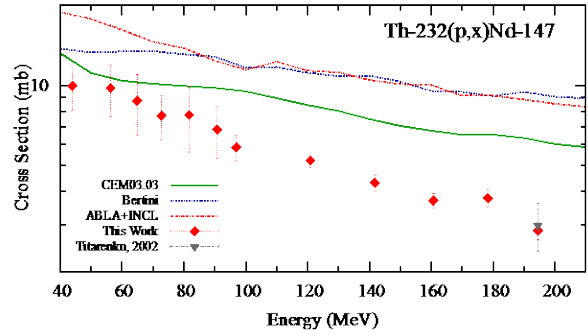

Fig. 5. Comparison of MCNP6 event generator predicted cross sections for the $^{232}$Th(p,x)$^{147}$Nd nuclear reaction from 40- to 200-MeV with experimental data.

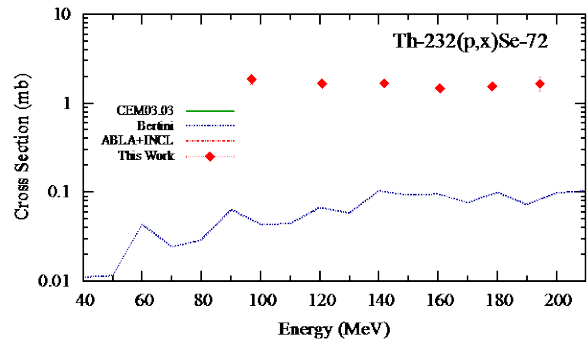

Fig. 6. Comparison of MCNP6 event generator predicted cross sections for the $^{232}$Th(p,x)$^{72}$Se nuclear reaction from 40- to 200-MeV with experimental data.

In the instance of the formation of $^{97}$Zr and $^{125}$Sb (Figs. 2 and 3), agreement of the predictions of the ABLA+INCL code with measured data exceeds that of CEM03.03 and Bertini, both of which under-predict the measured data by as much as a factor of 1.5. In contrast, $^{112}$Pd and $^{147}$Nd are best predicted by CEM03.03 (Figs. 4 and 5). The event generators studied here uniformly under-predict the measured $^{72}$Se cross section, with only the Bertini event generator approaching an order of magnitude of measured values (Fig. 6) and CEM03.03 predicting values more than an order of magnitude lower than these in turn.

No single event generator's predictions matched the measured values of an individual isotope across the measured energy range for all five isotopes measured. However, agreement between measured data and the event generators' predictions is within a factor of two across all energies measured and for all isotopes except $^{72}$Se, whose measured cross section is approximately a factor of 3 lower than the predictions of the Bertini code above 160 MeV incident proton energy.

Continued analysis of spectroscopic data is expected to produce additional cross sections of interest to event generator evaluation.

## ACKNOWLEDGEMENTS

We are grateful for technical assistance from LANL C-NR, C-IIAC, AOT-OPS, and LANSCE-NS groups' staff. This study was carried out under the auspices of the National Nuclear Security Administration of the U.S. Department of Energy at Los Alamos National Laboratory under Contract No. DE-AC52-06NA253996 with partial funding by the US DOE Office of Science via an award from The Isotope Development and Production for Research and Applications subprogram in the Office of Nuclear Physics.